# Characterization of LEO Satellites With All-Sky Photometric Signatures

**Harrison Krantz*, Eric C. Pearce*, and Adam Block***

*Steward Observatory, University of Arizona*

**Abstract**

We present novel techniques and methodology for unresolved photometric characterization of low-Earth Orbit (LEO) satellites. With the *Pomenis LEO Satellite Photometric Survey* our team has made over 14,000 observations of Starlink and OneWeb satellites to measure their apparent brightness. From the apparent brightness of each satellite, we calculate a new metric: the effective albedo, which quantifies the specularity of the reflecting satellite. Unlike stellar magnitude units, the effective albedo accounts for apparent range and phase angle and enables direct comparison of different satellites. Mapping the effective albedo from multiple observations across the sky produces an all-sky photometric signature which is distinct for each population of satellites, including the various sub-models of Starlink satellites.

Space Situational Awareness (SSA) practitioners can use all-sky photometric signatures to differentiate populations of satellites, compare their reflection characteristics, identify unknown satellites, and find anomalous members. To test the efficacy of all-sky signatures for satellite identification, we applied a machine learning classifier algorithm which correctly identified the majority of satellites based solely on the effective albedo metric and with as few as one observation per individual satellite. Our new method of LEO satellite photometric characterization requires no prior knowledge of the satellite's properties and is readily scalable to large numbers of satellites such as those expected with developing communications mega-constellations.

## 1. INTRODUCTION

The rapidly increasing numbers of low-Earth orbit (LEO) satellites pose significant challenges for Space Situational Awareness (SSA). Already we experience difficulty in space traffic management with increasing numbers of conjunctions and complications in identifying satellites after simultaneous deployments of multiple satellites [1]. Not yet fully addressed is the challenge of characterizing large numbers of satellites and fielding for anomalous members.

With development of low-cost launches and space capabilities around the world, we cannot expect all operators to be open or cooperative regarding information about their payloads. Between simultaneous deployment of potentially hundreds of satellites, and communications mega-constellations each consisting of thousands of individual satellites, it is not unthinkable that a single anomalous member could be lost amongst the crowd. Traditional means of singular observation and characterization are untenable when there are tens of thousands of individual satellites to investigate.

Numerous unresolved photometric and spectroscopic techniques for observing and characterizing space objects exist, but largely for satellites and debris in higher orbits such as geosynchronous (GEO) orbits. The same techniques are difficult or impossible to apply to LEO satellites. Satellites in LEO move much faster across the sky offering only short windows of approximately ten minutes to observe an individual satellite. Very few large or even small telescopes are capable of tracking LEO satellites with enough accuracy to keep the target satellite in view of the science sensor. LEO satellites are much closer in range, which requires a different perspective for interpreting changes in phase angle and overall Sun-satellite-observer geometry than with more distant GEO objects.

Due to the abundance and reliability of radar observations of LEO satellites in recent history, very little effort has gone to developing unresolved photometric characterization techniques for LEO. However, after the first Starlink satellites launched in 2019, astronomers and star gazers alike noted their bright nature and voiced concerns regarding the future of the night sky with thousands of similarly bright satellites. Suddenly the community called for photometric measurements of these satellites to determine their brightness and potential impact on astronomy [2][3].

## 2. POMENIS LEO SATELLITE PHOTOMETRIC SURVEY

For our study of mega-constellations and their impact on astronomy we created and executed an automated photometric survey to observe LEO satellites, specifically Starlink and OneWeb satellites [4][5]. Satellite brightness varies significantly with the Sun-satellite-observer geometry and in more ways than simple one-dimensional range and phase angle relations describe. To fully capture the range of brightness a satellite exhibits requires observing it across the entire range of possible geometries. For mega-constellation satellites, this includes the entire sky. We designed our survey to observe satellites at points across the entire sky to characterize the as-observed brightness of constellation satellites and directly assess the impact to astronomy as seen by an observer.

This observational survey utilizes low-cost equipment, specifically Pomenis, a system we previously built specially for SSA observations and studies [6]. Pomenis is an automated small-aperture wide-field of view system which lives in a custom-built mobile enclosure. The unique suite of capabilities is particularly suited for observing fast-moving LEO satellites.

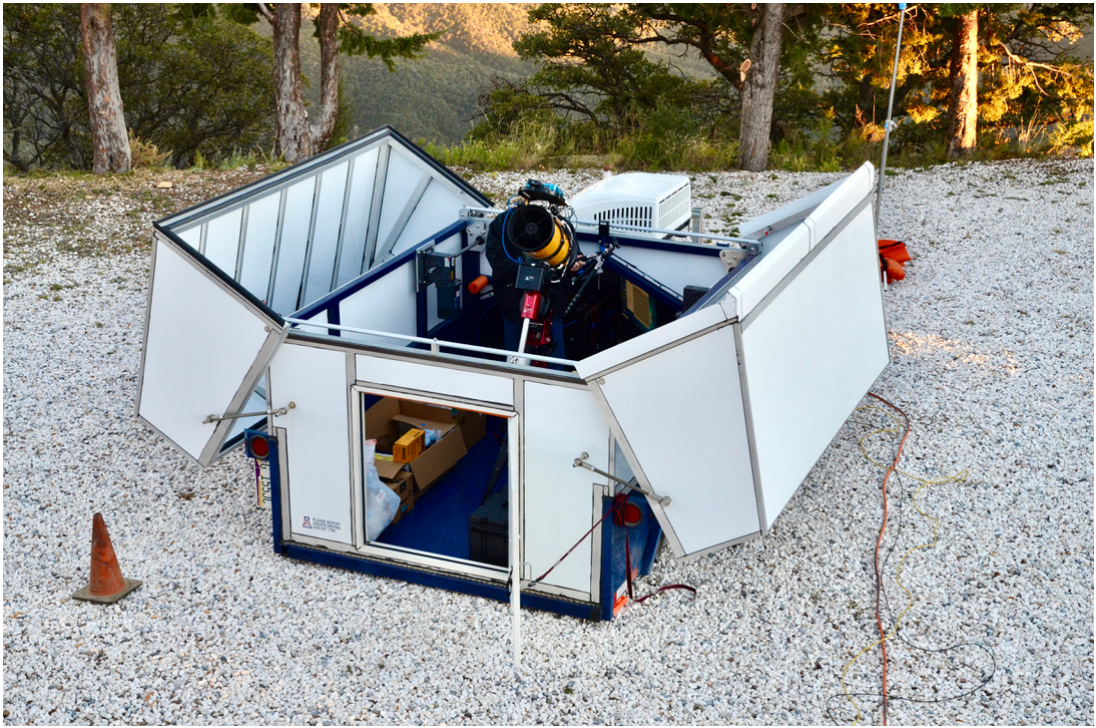

*Figure 1. Pomenis and the unique portable trailer-mounted enclosure.*

For the automated survey of mega-constellation LEO satellites we created a set of custom software programs which utilize the *Astropy* [7] and *Skyfield* [8] Python libraries. This includes software which determines which satellites are visible and intelligently creates a nightly schedule to observe as many satellites as possible while also prioritizing certain satellites with a weighting scheme to meet project goals.

To observe each individual satellite, we developed a "Wait & Catch" technique. In this technique, the telescope slews to specific coordinates on sky where we expect the satellite to be, waits for the satellite to arrive, and then records an image at the exact moment the satellite is passing through the field of view. This produces an image with the satellite streaked across the field and stationary un-streaked background stars. With the large field of view and a short exposure time, we capture the whole satellite streak within the image frame. Capturing the whole streak simplifies the photometry and enables unambiguous astrometric measurement of the satellite position on sky.

We created a semi-automated data processing pipeline to process the images. This process includes reducing the images with background subtraction and extraction of the satellite streak for photometry. Our software sums all pixel counts within the streak to determine the total brightness of the satellite. Since the whole satellite streak is within the image, the exposure time for the streak and background stars is the same, and the sum of the streak can be treated as a single measurement for relative photometry. To subtract background stars which overlap with the satellite streak, the survey records a "clean" image of the sky for every satellite observation. During photometry, our software sums the same pixels in the clean image capturing any background stars, and then subtracts this from the satellite streak. We utilize *Astrometry.net* [9] to plate solve the images and *The Photometry Pipeline* [10] to determine a reference photometric zero point from background stars. Our software utilizes the plate solution to determine the satellite's astrometric on-sky position measured at the streak centroid. All the information produced by the processing pipeline and otherwise associated with each observation is compiled into a *MySQL* database for correlation and easy querying.

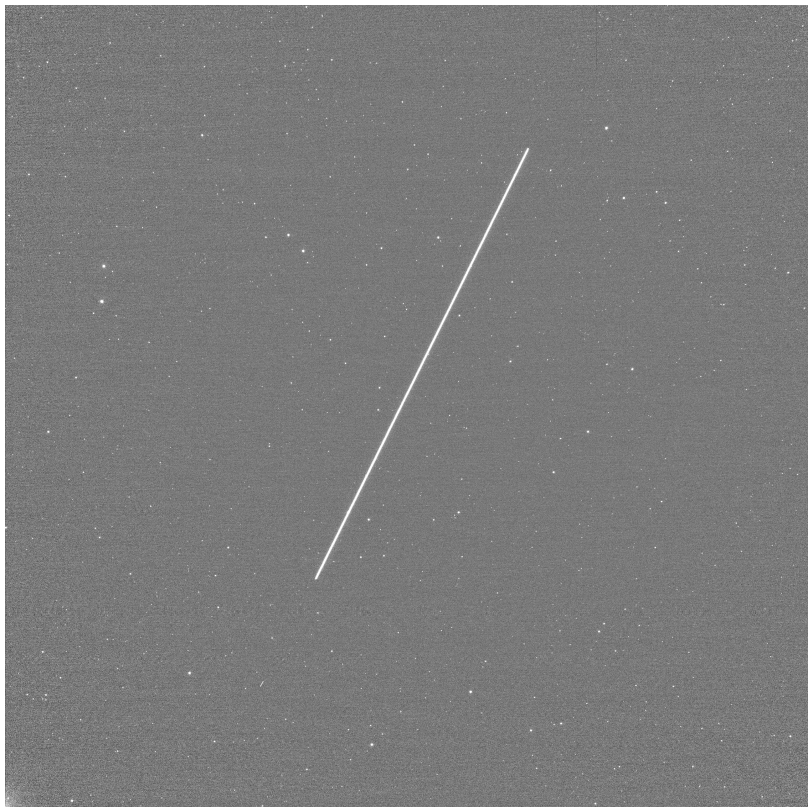

*Figure 2.* *An image of a bright satellite streak recorded by Pomenis.*

Over the previous two years, Pomenis has observed over 2,000 individual LEO satellites and produced over 14,000 photometric and astrometric measurements (as of June 1, 2022). We operate Pomenis and run the automated survey every twilight period barring extenuating circumstance.

*Table 1.* *The box-score for the Pomenis LEO Satellite Photometric Survey as of June 1, 2022.*

| Satellite Population | Number Observed | Number of Observations |
|---|---|---|
| *Standard Starlink* | 479 | 4759 |
| *Visor Starlink* | 985 | 4987 |
| *Laser Starlink* | 273 | 623 |
| *OneWeb* | 320 | 3623 |

## 2.1 SATELLITE MODELS

For the Pomenis LEO Satellite Photometric Survey, we observe Starlink and OneWeb satellites, which are the two currently active mega-constellations in LEO.

Over the previous two years of development, SpaceX modified the design of Starlink satellites multiple times, with three substantially different designs forming three distinct populations.

- The *Standard Starlink* population includes the first Starlink satellites launched in 2019 and those launched before mid-2020 when SpaceX modified the design to add visors [11]. We do not include the one-off prototype "DarkSat" in this population.

- The *Visor Starlink* population includes all the Starlink satellites which have underhanging shade visors; launched from mid-2020 to late-2021. SpaceX added these visors to reduce the apparent brightness of the satellites as seen by observers on the ground [11].

- The *Laser Starlink* population includes all Starlink satellites which have the external laser module; launched from late-2021 to present. The external laser module is for inter-communication between Starlink satellites. These satellites do not include shade visors [12].

At the time of writing, we are not aware of any substantial differences between any of the OneWeb satellites on orbit [13]. For analysis we treat all the OneWeb satellites as one population.

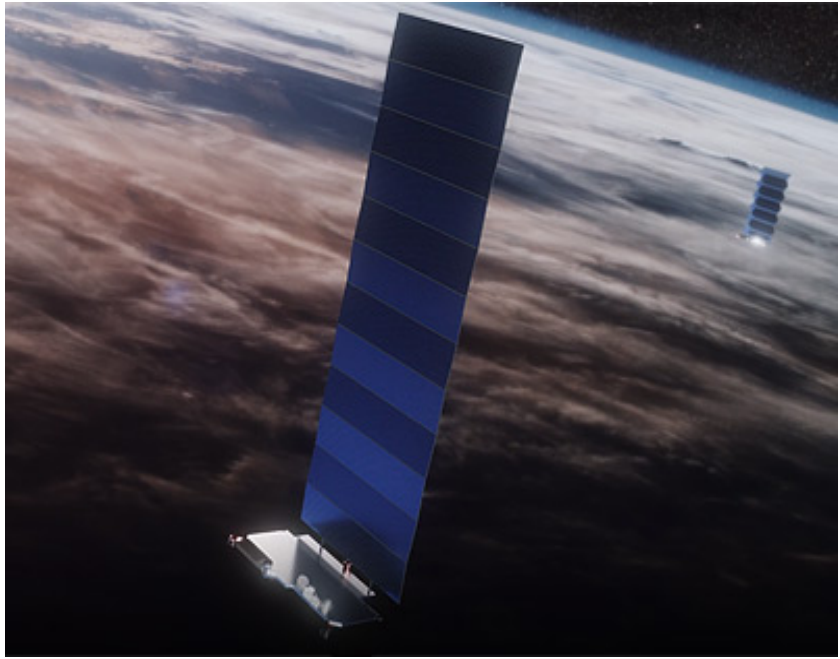

*Figure 3. A rendering of a Starlink satellite with solar array extended [11].*

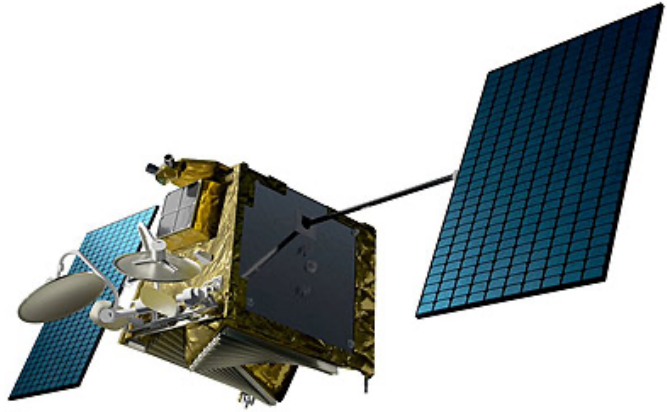

*Figure 4. A rendering of a OneWeb satellite showing the box-shaped satellite bus and pair of extended solar panel arrays [13].*

### 3. ALL-SKY PHOTOMETRIC SIGNATURES

To visualize the correlation of satellite brightness and Sun-satellite-observer geometry, we plot each individual photometric measurement at the position of observation on a map of the sky, with the color of each point representing the measured satellite brightness (**Figure 5**). However, this immediate all-sky plot does not show more than a scatter of seemingly uncorrelated values. This is because the entire collection of measurements covers a broad range of times, dates, and seasons, during which the Sun is not in the same location for each measurement. The Sun is also plotted below the horizon as a large yellow dot, which appears as a yellow swath along the edges of **Figure 5**.

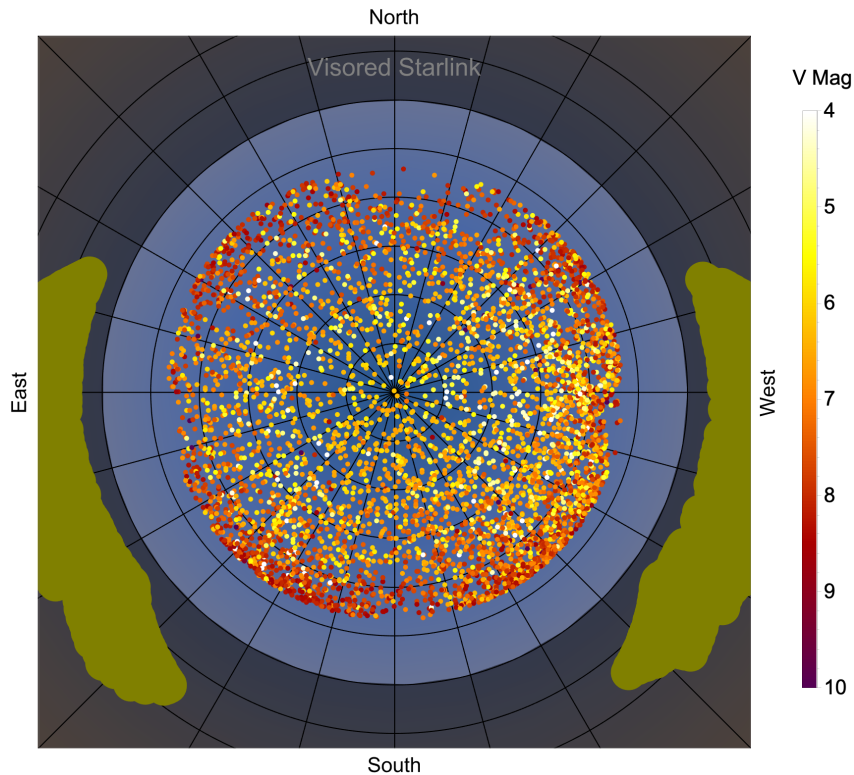

*Figure 5. All-sky plot showing the measured brightness of Visor Starlink satellites at the position on sky at the time of observation. The position of the Sun for each observation is also shown as a large yellow dot below the horizon.*

To produce a more useful all-sky plot, we rotate the plotted position of each measurement in azimuth around the zenith point such that the Sun's azimuth is the same for every measurement. This modification does not affect the measurement itself because the atmosphere and Sun-satellite-observer geometry is radially symmetric around the zenith point. There is risk that the satellite pose is not consistent relative to the Sun or direction of orbit; however without specific knowledge of the pose, we assume that for active satellites, pose is consistent.

We cannot do a similar modification to line up the below-horizon elevation of the Sun because changing the apparent elevation of each measurement changes the airmass and reflecting Sun-satellite-observer geometry. However, during the twilight period, when LEO satellites are visible, the Sun moves only a small amount in elevation and we do not expect a large change in the reflecting geometry or resulting all-sky pattern. To investigate this we created multiple all-sky plots and limited the points included in each plot to short ranges of Sun elevation. Comparing the plots we did not find any obvious changes in the apparent reflection pattern with small changes in the Sun's elevation.

Plotting the all-sky plot with rotated points (**Figure 6**) shows a clear pattern of apparent brightness across the sky.

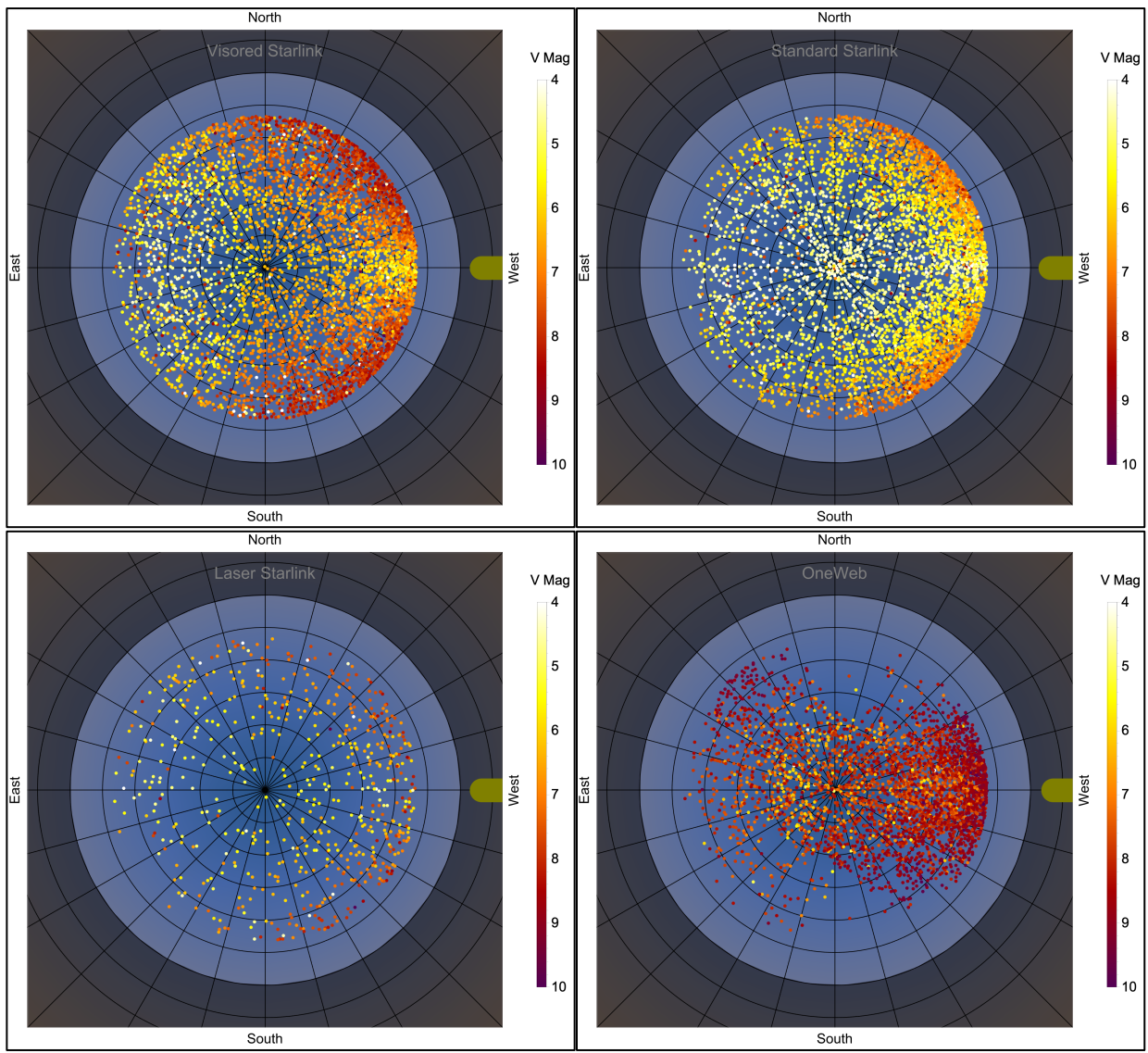

***Figure 6.*** *The rotated all-sky plots showing the pattern of apparent brightness across the sky for each population of satellites.*

# 4. EFFECTIVE ALBEDO METRIC

As a product of performing photometry relative to background stars we measure the apparent brightness of each satellite in stellar magnitude units. However, stellar magnitudes are a terrible metric for evaluating the reflection properties of space objects. Stellar magnitudes do not account for apparent phase angle, range, or other parameters which determine the quantity of light observed from a reflecting body.

One immediate approach inspired by the apparent brightness all-sky plots is to attempt fitting the data to some form of reflection model or bidirectional reflectance distribution function (BRDF). However, doing so is fraught with complications and perils, and may be entirely futile without prior knowledge of the design, shape, size, or materials of the space object.

In the interest of simple and universal analysis, we instead focus on a universal parameter which broadly describes the reflecting properties of a space object: its albedo.

The simplest and most commonly used method for calculating the apparent brightness of a reflecting space object is the diffuse sphere model. In this model we treat the space object as a gray uniform sphere with perfect Lambertian reflectance. Then accounting for the object's range $R$, phase angle $\theta$, size $A$, and albedo $\rho$ we calculate the expected apparent brightness $M_v$ in **Equation 1** [14].

$$M_v = M_{Sun} - 2.5 \log\left(\frac{2}{3\pi^2} A\rho[(\pi - \theta)\cos\theta + \sin\theta]\right) + 5 \log R \tag{1}$$

The diffuse sphere model is adequate for approximating the brightness of space objects but does not accurately predict the brightness of many satellites in many geometries because satellites are not uniform diffuse spheres. Satellites are complex objects with many discrete surfaces and complex shapes. The combination of multiple reflections from diffuse and specular surfaces produces the summation of reflected light which we observe as the apparent brightness. Even slight changes in illumination geometry or satellite orientation can yield large changes in apparent brightness, which the diffuse sphere model does not include.

Here we utilize the diffuse sphere model in a different way. We use it as a known benchmark. We know how an ideal uniform diffuse sphere will reflect light and compare our observed satellite to it. After inverting **Equation 1** to create **Equation 2**, we input our measured apparent brightness $M_v$, range $R$, phase angle $\theta$, assign a default unit size for $A$, and calculate the albedo. This produces what we call the *effective albedo*, $\rho_{eff}$. It is the albedo needed for an ideal diffuse sphere of unit size to produce the observed brightness in the observed geometry.

$$\rho_{eff} = \frac{3\pi^2 R^2 10^{\frac{M_{Sun} - M_v}{2.5}}}{2[(\pi - \theta)\cos\theta + \sin\theta]} \tag{2}$$

The effective albedo is not a real measurement of albedo and should not be interpreted as such. The utility in this metric is the relative change in value across different reflecting geometries. A higher than baseline effective albedo indicates a more specular reflection which is directly inherent to the shape, design, and materials of the satellite.

For the same photometric data shown in **Figure 6**, we computed the effective albedo for each measurement and created all-sky plots (**Figure 7**). These all-sky plots show the relative reflectivity in different geometries and reveal where in the sky the satellite appears more or less specular. Each population of satellites exhibits a distinctive pattern, and because we accounted for range and phase angle, we can directly compare reflection patterns of different satellite populations.

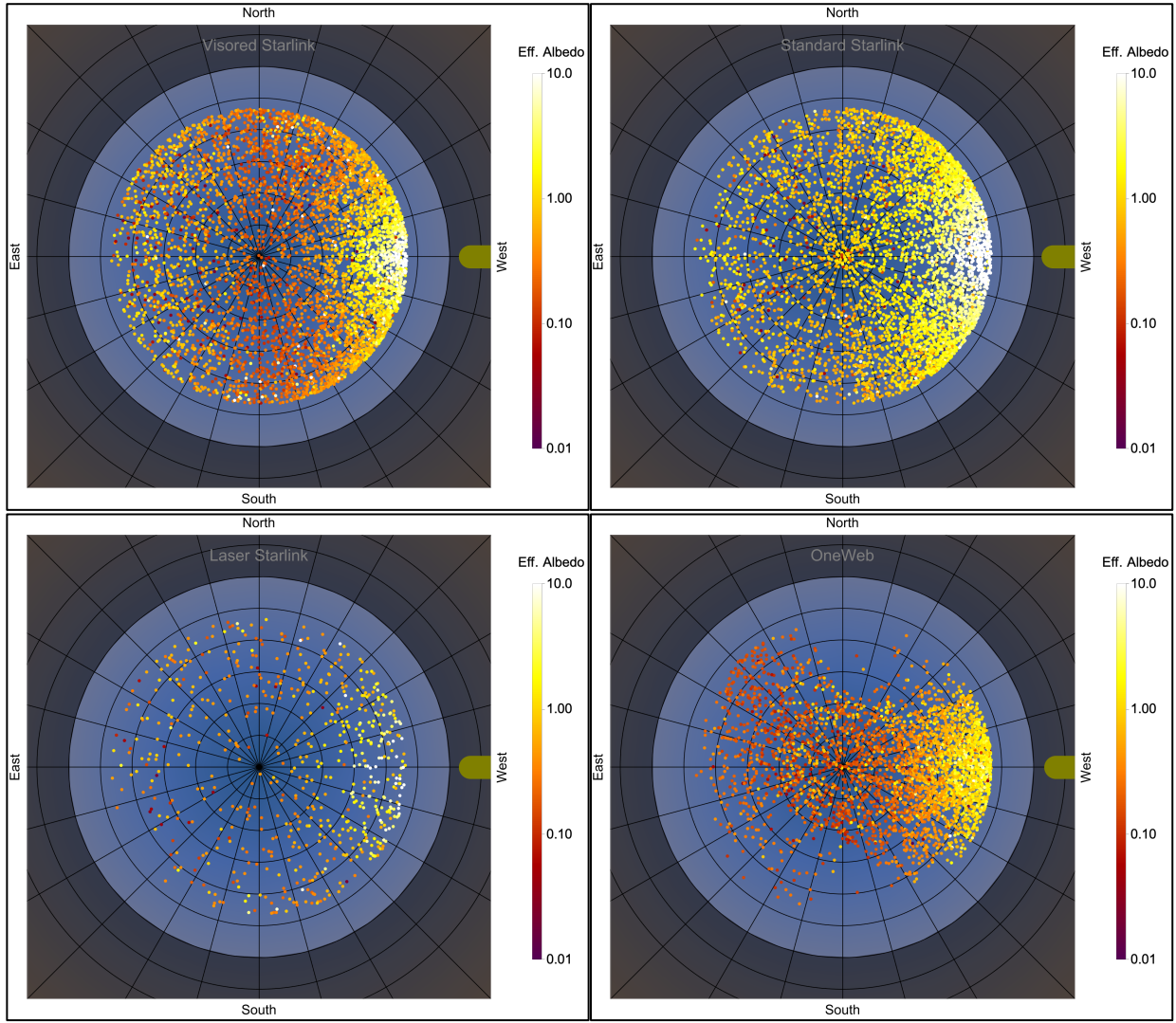

***Figure 7.*** *All-sky plots showing the pattern of effective albedo across the sky for each population of satellites.*

## 5. IDENTIFICATION BY MACHINE LEARNING CLASSIFIER

The effective albedo all-sky plots show distinctive reflection signatures for each population of satellites. We can leverage this to identify the corresponding population of an unknown satellite by comparing it to the known populations. To do this, we utilized a gradient boosted decision trees classifier algorithm [15] with the existing data set as both a training set and test set. We ran the classifier separately for every individual satellite in the data set as a test satellite and compiled the results to determine the overall efficacy of this method to identify satellites. For each individual test satellite, we removed its measurements from the complete data set and used this reduced set to train the classifier algorithm. The training data includes the elevation angle, modified azimuth angle, effective albedo, and known satellite population for each data point. After learning on the training data, the classifier attempted to identify the test satellite and determine which population it belongs to.

Overall the classifier is highly accurate and correctly identified the majority of satellites for each population, except the *Laser Starlink* population.

*Table 2. The classifier accuracy for each population of satellites.*

| Satellite Population | Number Correct | Number Incorrect | Accuracy |
|---|---|---|---|
| Standard Starlink | 462 | 17 | 96.5% |
| Visor Starlink | 847 | 138 | 86.0% |
| Laser Starlink | 0 | 273 | 0.0% |
| OneWeb | 287 | 33 | 89.7% |

None of the *Laser Starlink* satellites were correctly identified with the classifier falsely tagging the majority as *Visor Starlink*, but only with low confidence levels. We conclude the classifier failed to correctly identify the *Laser Starlink* satellites due to a combination of reasons:

- The majority of *Laser Starlink* satellites have only one or two observations in the data set, severely limiting the information available to the classifier (this impact is further addressed in the next paragraph).
- The total number of *Laser Starlink* observations is much less than the other populations, and the all-sky plot is not as complete. In addition to the *Laser Starlink* population being less characterized, the imbalance in training data also produces bias in the machine learning algorithm we used.
- The *Laser Starlink* design is very similar to the also visor-less *Standard Starlink* design and may be difficult to differentiate. Though we note that only a minority of satellites were falsely identified at *Standard Starlink*.

The accuracy of the classifier is heavily dependent on the number of observations for an individual satellite. More observations of an individual satellite provide more data to the classifier and undoubtedly impacts the ability to identify it. Additionally, some observations are more informative than others. The various populations of satellites look similar in some geometries but different in others. For the purpose of identifying satellites, future strategic observations could target specific geometries to better differentiate satellite populations.

To determine how many observations are necessary to produce an accurate identification, we ran the classifier again but only provided $n$ (*1, 2, 3…*) observations for each test satellite. **Figure 8** shows that the accuracy of the classifier increases when provided more observations for each individual test satellite. With only a single observation, the classifier does correctly identify a majority of satellites for the three populations. Providing more observations per satellite increases the accuracy, and less than 5 are necessary to reach greater than 80% accuracy.

Note that only a small fraction of the satellites in the data set have more than 10 observations, which significantly reduces sample size for calculating the overall accuracy and produces bumps in the curve.

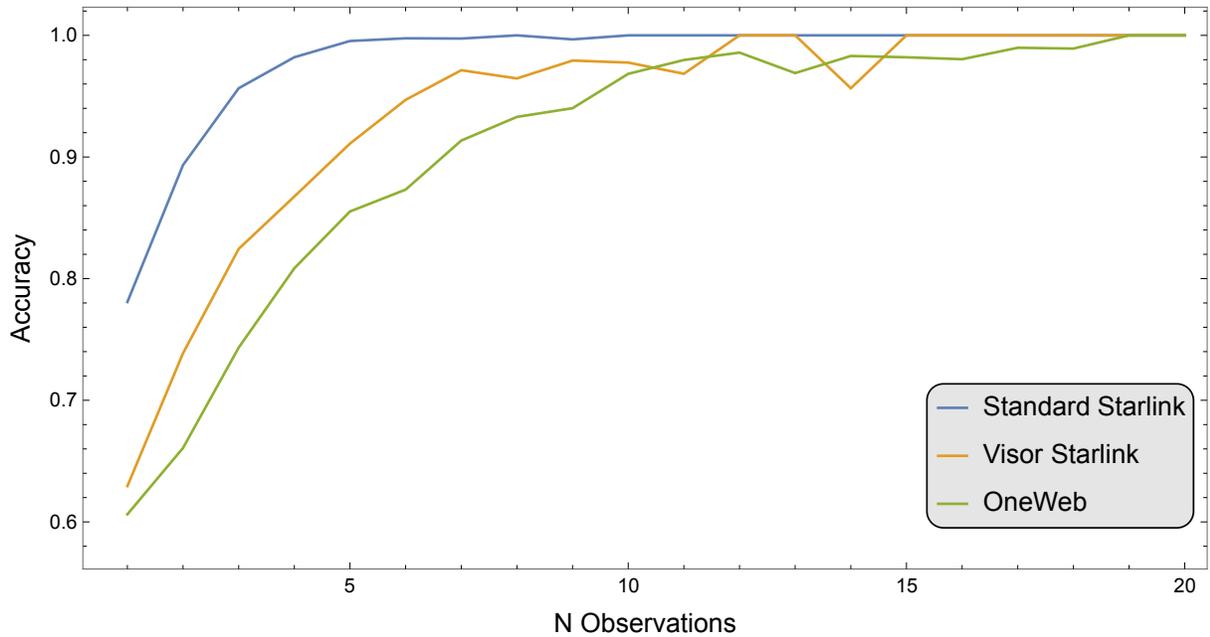

***Figure 8.*** *This plot shows the accuracy of the classifier as a function of the number of observations provided for each individual test satellite. As more observations are provided, the classifier becomes more accurate and correctly identifies more satellites.*

## 6.   CONCLUSIONS

Over the previous two years, the *Pomenis LEO Satellite Photometric Survey* has produced over 14,000 photometric and astrometric measurements of over 2,000 individual LEO satellites. Creating all-sky plots of the observed brightness shows distinct patterns for each population of satellites. Due to limitations in characterizing reflecting bodies with stellar magnitudes, we created a new metric: the effective albedo, which describes the specularity of an individual satellite in a specific Sun-satellite-observer geometry. Because the effective albedo accounts for the apparent range and phase angle, we can directly compare measurements in different geometries and for different satellite populations.

As one example of how to leverage the information produced through all-sky characterization and the effective albedo metric, we attempted to identify satellites with a machine learning classifier. Overall the classifier algorithm is highly accurate with one exception: the classifier failed to correctly identify the *Laser Starlink* satellites due to a myriad of reasons, stemming from a reduced data set compared to the other populations of satellites. Otherwise, the classifier correctly identified the majority of satellites, even with only a single observation provided for each individual test satellite. Fewer than 5 observations per individual satellites are needed to exceed 80% accuracy.

Going forward, several possible improvements to our methodology can significantly expand the value of the all-sky photometric signatures. Multiple telescopes could be employed to quickly observe and characterize newly launched satellites. Observations in additional photometric filters could produce accompanying all-sky signatures in different colors, enabling all-sky color index plots. Better knowledge of satellite populations and subtle variations in design will improve the training set for identification. More sophisticated machine learning algorithms can address the specific task of satellite identification from all-sky signatures and better handle imbalanced data sets. Anomaly detection algorithms can test and flag individual satellites which do not appear like the parent population.

There are many possibilities for how to utilize the all-sky photometric signatures for characterizing populations of satellites. With over a dozen satellite mega-constellations expected in the next decade comprised of as many as one hundred thousand individual satellites, monitoring and characterizing large numbers of LEO satellites is going to become a significant challenge.

Our novel method of characterization is readily scalable to vast numbers of satellites and is simple to implement, requiring only low-cost small telescopes. An autonomous array of telescopes could survey the sky, observing hundreds to thousands of individual satellites every night. The growing number of photometric measurements, combined with ground knowledge, would create baseline all-sky photometric signatures for future comparison of new satellites and outlier identification. Application of machine learning techniques could automate much of the analysis and even allow for automatic differentiation of satellites and creation of all-sky photometric signatures without ground knowledge.